\documentstyle[twoside,fleqn,jhuwkshp,psfig]{article}

\begin{document}

\title{900 KS EXPOSURE OF NGC\,3783 WITH {\it CHANDRA}/HETGS: \\
NO ACCRETION DISK LINES}

\author{S. Kaspi \thanks{In collaboration with W. N. Brandt, Ian
M. George, Hagai Netzer, D. Michael Crenshaw, Jack R. Gabel, Frederick
W. Hamann, Mary Elizabeth Kaiser, Anuradha Koratkar, Steven B. Kraemer,
Gerard A. Kriss, Smita Mathur, Richard F. Mushotzky, Kirpal Nandra,
Bradley M. Peterson, Joseph C. Shields, T. J. Turner, and Wei Zheng.}
\address{Department of Astronomy and Astrophysics, 525 Davey Laboratory,
The Pennsylvania State University, University Park, PA 16802, USA.}}

\begin{abstract}
We present preliminary results from a 900 ks exposure of NGC\,3783 with
the {\it Chandra}/HETGS. This is the best combination of signal-to-noise
and resolution ever obtained for an Active Galactic Nucleus (AGN). We
resolve the narrow Fe\,K$\alpha$ line to have FWHM of $1720\pm360$
km\,s$^{-1}$ which, under the simple assumption of virialized system,
suggest that this narrow line is emitted between the broad line region
and the narrow line region. We do not detect any broad component for the
Fe\,K$\alpha$ line though such component was observed in previous {\it
ASCA} observations of this object. Our results suggest an evolution in
the broad Fe\,K$\alpha$ line that took place from the observations in
1996 to the observations in 2000/2001.

The high resolution X-ray spectrum of NGC 3783 shows more than a hundred
absorption lines and several dozen emission lines from the H-like and
He-like ions of N, O, Ne, Mg, Al, Si, and S as well as from Fe\,{\sc
xvii} to Fe\,{\sc xxiv} L-shell transitions. All these features can be
modeled by a multi-component, outflowing, photoionized absorber. There
is no evidence (or need in the model) for soft X-ray emission lines from
a relativistic accretion disk as have been proposed to be seen in a few
other AGNs.
\end{abstract}

\maketitle

\section{Introduction}

NGC\,3783 is a bright Seyfert~1 galaxy in which its X-ray spectrum shows
some of the strongest absorption features around 0.7--1.5 keV. Studies
by {\it ROSAT} \cite{T93} and {\it ASCA} (e.g., \cite{G98} and references
therein) have shown that the 2--10 keV continuum is fitted by a power law
with photon index $\Gamma \approx 1.7$--1.8, the 2--10 keV flux varies
in the range $\approx(4$--$9)\times 10^{-11}$~ergs\,cm$^{-2}$\,s$^{-1}$,
and its mean X-ray luminosity is $\approx 1.5\times 10^{43}$~ergs s$^{-1}$
(for $H_0=70$~km\,s$^{-1}$\,Mpc$^{-1}$ and $q_0=0.5$). The absorption
features are attributed to O\,{\sc vii} and O\,{\sc viii} edges and
their models indicate a column density of ionized gas of $\approx 2
\times 10^{22}$~cm$^{-2}$.

On 2000 January 21 NGC\,3783 was observed with the High-Energy
Transmission Grating Spectrometer (HETGS) on the {\em Chandra X-ray
Observatory\/} with the Advanced CCD Imaging Spectrometer (ACIS) as the
detector. This spectrum shows several dozen absorption lines and a few
emission lines from the H-like and He-like ions of O, Ne, Mg, Si, and S
as well as from Fe\,{\sc xvii}--Fe\,{\sc xxiii} L-shell transitions. The
absorption lines are blueshifted relative to the systemic velocity by
$\approx -610$ km\,s$^{-1}$ while the emission lines are consistent with
being at the systemic velocity \cite{K2000,K2001}. High-resolution UV
spectra of NGC\,3783 taken with {\it HST} show intrinsic absorption
features due to C\,{\sc iv}, N\,{\sc v} and H\,{\sc i} (e.g.,
\cite{KC2001} and references therein). Currently there are three known
absorption systems in the UV at radial velocities of $\approx -$560,
$-$720, and $-$1400 km\,s$^{-1}$ (blueshifted) relative to the optical
redshift. The strength of the absorption in the UV and in the X-ray bands,
is found to be variable over time scales of months to years.

The above characteristics have made NGC\,3783 to be a perfect target for
an intensive study of the ionized gas in its nuclear environment. This
multiwavelength monitoring study was carried out during 2001
February--June by {\it Chandra}/HETGS, {\it RXTE}, {\it HST}/STIS,
{\it FUSE}, and a ground based observatory. The {\it Chandra}/HETGS
observations consist of five observations each of $\approx$\,170
ks and adding the initial 56 ks the total exposure time (ONTIME)
is 900.1~ks (total of good time interval corrected for detector dead
time is 888.7~ks). In this contribution we will present the mean 900 ks
{\it Chandra}/HETGS X-ray spectrum of NGC\,3783. We will focus on the
Fe\,K$\alpha$ region where a broad line, modeled as an emission line from
accretion disk, was detected in {\it ASCA} observations from 1996. The
detailed {\it Chandra}/HETGS spectra and analysis will be presented in
Kaspi et al., in preparation (mean X-ray spectrum), I. M. George et al.,
in preparation (X-ray variability), and H. Netzer et al., in preparation
(X-ray spectra models). {\it HST}/STIS and {\it FUSE} observations
will be presented in Crenshaw et al., in preparation, and Gabel et al.,
in preparation, respectively.

\section{900 ks mean X-ray spectrum}

\begin{figure*} 
{\psfig{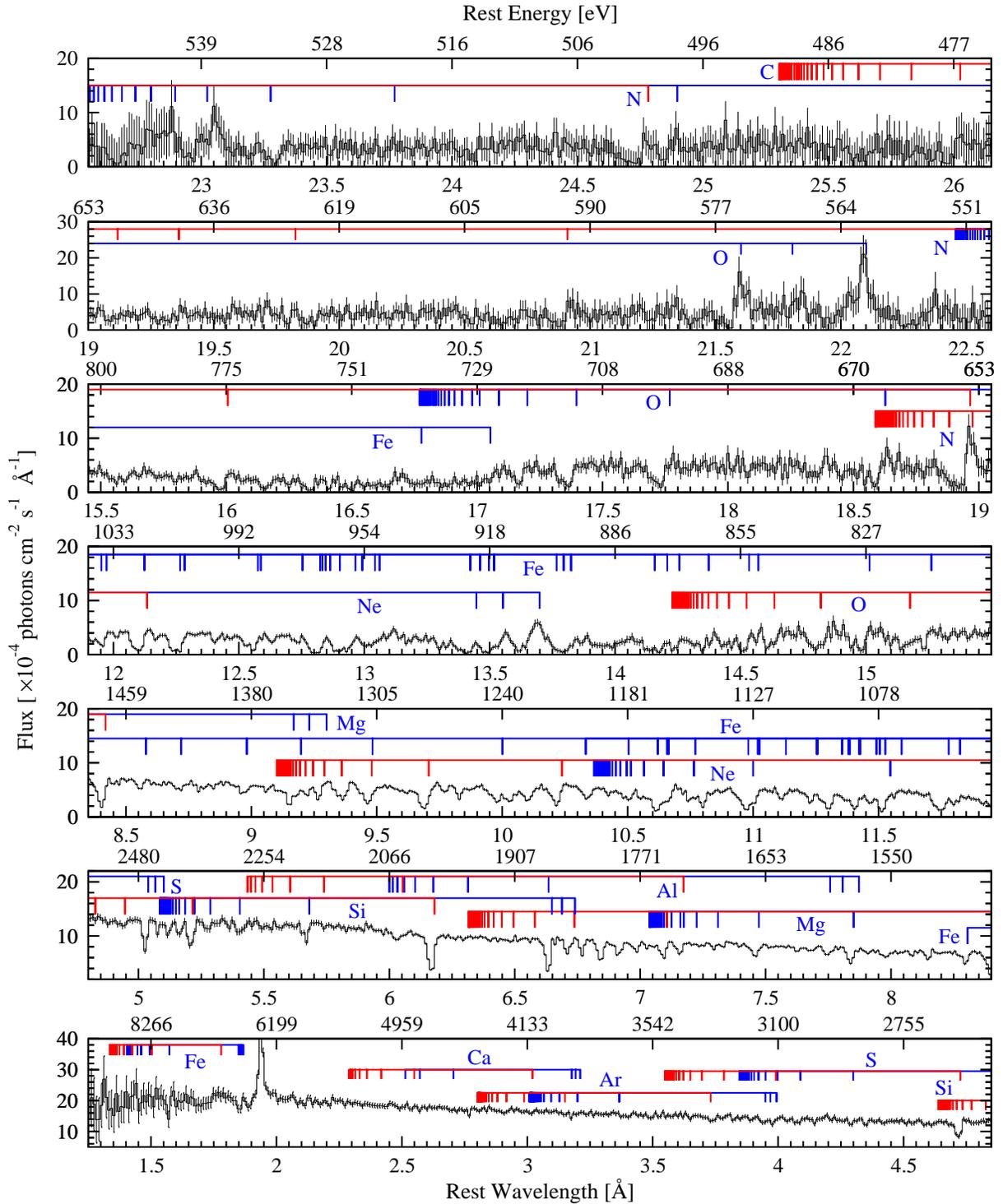}}
\vglue -0.7cm
\caption{Combined MEG and HEG 900 ks spectrum binned to
0.01 \AA . Each data point has an error bar representing its
uncertainty. The H-like and He-like lines of the identified ions are
marked in red and blue, respectively. For each ion the theoretically
expected lines are plotted up to the ion's edge (not all lines are
identified in the data). The ions' lines are marked at their expected
wavelength in the rest frame of NGC3783, and the blueshift of the
absorption lines is noticeable.
}\label{specfig}
\end{figure*}

All {\it Chandra}/HETGS observations were reduced uniformly and
in the standard way using the {\it Chandra} Interactive Analysis of
Observations (CIAO) software (Version 2.1.2), and its Calibration Database
(Version~2.6). The mean 900 ks spectrum is shown in Fig.~\ref{specfig}.
About 150 line features can be identified in the spectrum; many of these
are blends of several lines. We identify absorption lines from H-like
and He-like ions of N, O, Ne, Mg, Al, Si, and S as well as absorption
from lower-ionization ions such as Si\,{\sc vii}--Si\,{\sc xii} and
S\,{\sc xii}--S\,{\sc xiv}. There are also many absorption lines from
iron ions; L-shell and M-shell lines of Fe\,{\sc xvii}--Fe\,{\sc xxiv}
as well as probable resonance lines of Fe\,{\sc xxv}. Absorptions by C,
Ar, and Ca are hinted, although these are not significant.

Focusing on the topic of this meeting, there are two regions were emission
lines from accretion onto black holes are expected: the Fe\,K$\alpha$
line at 6.4 keV which will be discussed in the next section, and
the soft X-ray emission lines from a relativistic accretion disk
recently suggested to be present in the narrow line Seyfert~1 galaxies
MCG\,{$-$}6-30-15 and Mrk\,766 using {\it XMM-Newton} observations
\cite{BR2001}. These are Ly$\alpha$ lines from O\,{\sc viii}, N\,{\sc
vii}, and C\,{\sc vi} which are broadened by relativistic effects in the
accretion disk and appear in the spectrum as broad saw-toothed features
around 17--35~\AA.  When binning the 900 ks X-ray spectrum of NGC\,3783
to {\it XMM}/RGS resolution, it has adequate signal-to-noise ratio up to
about 27~\AA. Examination of this part of the spectrum reveals no such
features as identified in MCG\,{$-$}6-30-15 and Mrk\,766. Preliminary
model for the data, based on the multi-component, outflowing, photoionized
absorber model presented in \cite{K2001}, fits the 900~ks spectrum over
the whole 0.5--10 keV band, with no need to invoke the relativistic
accretion disk lines.

\section{The Fe\,K$\alpha$ line region}

\begin{figure}[t]
\vspace{10pt}
\centerline{\psfig{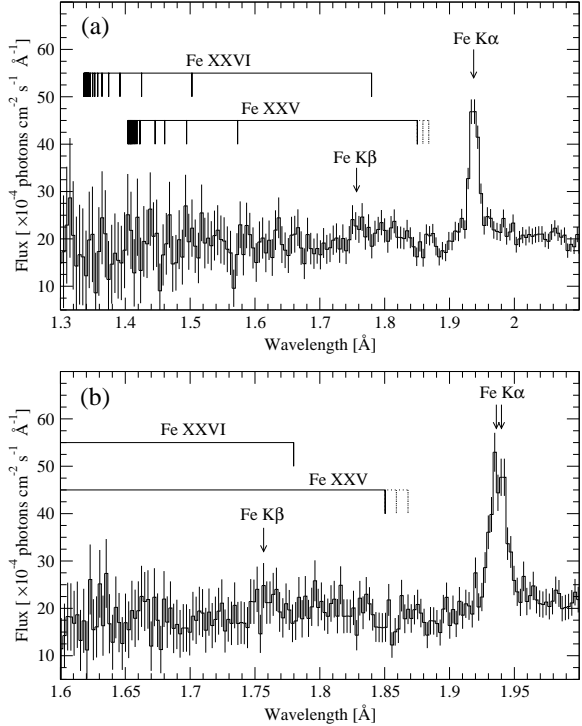}}
\vglue -0.9cm
\caption{HEG spectrum around the Fe\,K$\alpha$ feature: (a) binned to
0.005~\AA\ and (b) binned to 0.0025~\AA. In (b) the two Fe\,K$\alpha$
lines are marked. Also shown are the theoretical wavelengths of Fe\,{\sc
xxv} and Fe\,{\sc xxvi} absorption lines as well as the forbidden and
intercombination lines of Fe\,{\sc xxv}.
}\label{feplot}
\vglue -0.5cm
\end{figure}

The high-resolution X-ray spectrum of NGC\,3783 shows a prominent narrow
Fe\,K$\alpha$ emission line. A Gaussian fitted to the 0.0025~\AA\
binned spectrum (Fig.~\ref{feplot}b) gives a central wavelength of
$1.9378\pm0.0010$~\AA\ ($6398.2\pm3.3$~eV) which is consistent with the
Fe\,K$\alpha$ line from Fe\,{\sc i} to Fe\,{\sc xi}. Interestingly,
the Fe\,K$\alpha$ line in Fig.~\ref{feplot}b shows two peaks (though
these are not significantly resolved) which are consistent with the two
expected Fe\,K$\alpha$ lines for Fe\,{\sc i}, K$\alpha_1$ at 1.936~\AA\
(6403.84 eV) and K$\alpha_2$ at 1.940~\AA\ (6390.84 eV), with a branching
ratio of 2:1. We fitted the HEG spectrum with two Gaussians fixed at the
wavelengths of Fe\,K$\alpha$ lines and with the same branching ratio. We
find the FWHM of the Gaussians to be $16.3^{+1.7}_{-1.5}$~m\AA\ and, when
taking into account the instrumental FWHM of 12~m\AA, we get a true FWHM
of $11.1\pm 2.3$~m\AA. This FWHM corresponds to $1720\pm360$ km\,s$^{-1}$
at the wavelength of the Fe\,K$\alpha$ line (fitting the data with only
one Gaussian yields a consistent result, $1860\pm340$ km\,s$^{-1}$).

If we assume the very simple assumption that the gas around the
central mass is moving as a virialized system than there is a simple
anticorrelation of the line width with the radial location of its
origin. Since the Broad lines in NGC\,3783 has line widths of order
4000 km\,s$^{-1}$ and the Narrow lines has line widths of order 500
km\,s$^{-1}$, the resolved line width of the narrow Fe\,K$\alpha$
line suggests an origin in between the BLR and the NLR. Within the
frame of AGNs' unified models this is consistent with an emission
from the torus. Indeed, the Fe\,K$\alpha$ narrow line equivalent width
($27.4\pm3.3$~m\AA\ = $90\pm11$~eV) and flux ($[5.26\pm0.63]\times10^{-5}\
{\rm photons\ cm^{-2}\ s^{-1}}$) are consistent with models predicting
emission from the torus (e.g., \cite{K94}).

The narrow Fe\,K$\alpha$ line seems to have a red wing extending to
$\approx 2$~\AA\ ($\approx 6.2$~keV). We tentatively identify this
red wing as the ``Compton shoulder'' produced by Compton scattering in
optically-thick cold matter which can be identified with the obscuring
torus. The shoulder extends from 6.2 to 6.4~keV, has a total flux of
$(8.6\pm2.7)\times10^{-6}\ {\rm photons\ cm^{-2}\ s^{-1}}$, and its
EW is $4.2\pm1.3$ m\AA . These numbers are in agreement with previous
models and observations of such a shoulder (e.g., \cite{I97}).

\begin{figure}[t]
\vspace{10pt}
\centerline{\psfig{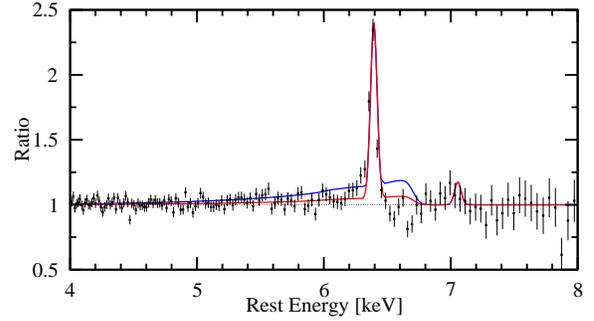}}
\vglue -0.9cm
\caption{The observed ratio of the data to the underlying continua near
the Fe\,K$\alpha$ line. Data are binned at 0.01~\AA\ to allow comparison
with Fig.~9 of \cite{K2001}. {\it Red line:} The model prediction for
a narrow Gaussian plus an {\em upper limit} for a 6.4~keV Schwarzschild
``disk line''. {\it Blue line:} Same model with its intensity as measured
from the 1996 {\it ASCA\/} observations \cite{G98,K2001}. Most data points
are below this model, suggesting that the intensity of the Fe~K$\alpha$
line has decreased since 1996.
}\label{femodel}
\vglue -0.4cm
\end{figure}

We have looked for a broad component for the Fe~K$\alpha$ line by fitting
an absorbed power-law continuum to the energy range 2.5--10~keV (excluding
the 5--7 keV range and several narrow absorption lines) and modeling the
Fe~K$\alpha$ line with a narrow Gaussian plus a ``disk-line'' component
for a Schwarzschild black hole \cite{F89}. This procedure is described
in detail in \cite{K2001}. We do not find a broad component in the 900
ks {\it Chandra}/HETGS spectrum (see Fig.~\ref{femodel}) and we are
only able to place an upper limit on its intensity to be $A_{\rm bl} <
4.2\times10^{-5}\ {\rm photons\ cm^{-2}\ s^{-1}}$ (EW $< 60$~eV).

In {\it ASCA} observations carried out in 1996 the Fe~K$\alpha$
line is modeled with a broad component with an intensity of
$(12\pm4)\times10^{-5}\ {\rm photons\ cm^{-2}\ s^{-1}}$ and a narrow
component with an intensity of $(4.2\pm1.6)\times10^{-5}\ {\rm photons\
cm^{-2}\ s^{-1}}$ \cite{G98,K2001}. While the narrow component flux
is consistent with the current measurement, the upper limit we set for
the broad component indicates that its flux decreased by a factor of at
least three between the 1996 observations and the 2000/2001 observations.
Other line models, which are still need to be constrained, will be presented in
Kaspi et al., in preparation.


\normalsize

\section*{ACKNOWLEDGEMENTS}

We gratefully acknowledge the financial support of CXC grant GO1-2103,
and NASA LTSA grant NAG~5-8107.

\end{document}